\documentstyle[aps,multicol,psfig]{revtex}  %...option for REVTEX 3.0 
\input epsf
\begin{document}
%--------------------------
%
%--------------------------
\newcommand{\shmd}{{E}}
\newcommand{\shex}{{t}}
\newcommand{\cpo}{{W}}
\newcommand{\fft}{\bf}
\newcommand{\wimpf}{\varpi}
%--------------------------
\draft
\title{
	Elasticity near the vulcanization transition}
\author{Horacio E.~Castillo and Paul M.~Goldbart}
\address{Department of Physics, University of Illinois at
Urbana-Champaign, 1110 West Green Street, Urbana, IL 61801-3080, USA} 
\date{December 3, 1997}
%\date{\today}
\maketitle
%--------------------------
%--------------------------
\begin{abstract}
Signatures of the vulcanization transition---amorphous solidification
induced by the random crosslinking of macromolecules---include the
random localization of a fraction of the particles and the emergence
of a nonzero static shear modulus. A semi-microscopic
statistical-mechanical theory is presented of the latter signature
that accounts for both thermal fluctuations and quenched disorder. It
is found (i)~that the shear modulus grows continuously from zero at
the transition, and does so with the classical exponent, i.e., with
the third power of the excess cross-link density and, quite
surprisingly, (ii)~that near the transition the external stresses do
not spoil the spherical symmetry of the localization clouds of the
particles.
\end{abstract}
%--------------------------
%--------------------------
%
% Possible Pacs numbers 
% (from 1996 Physics and Astronomy Classification Scheme (PACS)
% at http://www.aip.org/pacs/pacscheme.html) :
% 05.70.Fh  Phase transitions: general aspects
% 61.43.-j  Disordered solids
% 61.43.Dq  Amorphous semiconductors, metals, and alloys
% 61.43.Er  Other amorphous solids
% 61.43.Hv  Fractals; macroscopic aggregates (including diffusion-limited
%           aggregates)
% 64.60.Ak  Renormalization-group, fractal, and percolation studies 
%           of phase transitions 
% 64.70.Dv  Solid-liquid transitions
% 82.70.Gg  Gels and sols
%
%--------------------------
%--------------------------
\begin{multicols}{2}
%--------------------------
%--------------------------
%--------------------------
\noindent
{\em Introduction\/}: When a sufficient density of randomly located
cross-links is imposed on a system of flexible linear macromolecules, an
equilibrium phase transition (known as the vulcanization transition)
occurs.  At this transition a liquid state is replaced by an amorphous
solid state.  This transition has two main equilibrium signatures: (i)~a
nonzero fraction of the monomers become localized around random mean
positions and with random localization lengths (structure); and
(ii)~the system, as a 
whole, acquires a nonzero static shear modulus (response). The former
signature has 
been discussed previously; the purpose of the present Letter is to address
the latter signature. Specifically, our aim is to develop a
statistical-mechanical theory of the elastic properties of the
amorphous solid state 
in the vicinity of the vulcanization transition.  This theory
incorporates both annealed (i.e.\ thermally equilibrating) and quenched
random (i.e.\ cross-link specifying) variables.  Its primary conclusions
are: (a)~that the amorphous solid [in the sense of
signature~(i)] state emerging at the vulcanization transition is indeed a
solid [in the sense of signature~(ii)]; (b)~that the shear modulus
vanishes continuously as the transition is approached, and does so with
the third power of the excess cross-link density (i.e.\ the amount by
which the cross-link density exceeds its critical value); and (c)~that
the shearing of the container associated with elastic deformations does
{\it not\/} lead to a shearing of the probability clouds associated with
the thermal fluctuations of localized particles about their mean 
positions.

The elastic properties of vulcanized matter and related
chemically-bonded systems, especially those near the amorphous
solidification transition, have received considerable attention to
date.  Notable approaches include the classical
ones~\cite{REF:DobGor}, in which it was argued that near the
transition the elastic entropy in the solid phase (and consequently
the static shear modulus $\shmd$) grow as the third power of the
excess cross-link density $\epsilon$, i.e.,
$\shmd\sim\epsilon^{\shex}$ with $\shex=3$.  Subsequently, it was
proposed that the amorphous solidification transition of polymer
systems be identified with a percolation
process~\cite{REF:PGDGbook,REF:PGDGperc}. Thus, the exponent $\shex$
was identified with the critical exponent $\mu$ for percolation of
conductivity (with $\mu\approx 2.0$ in $3$ spatial
dimensions). Subsequently it was observed that the elasticity
percolation exponent for a random network is substantially higher than
$\mu$ when the forces are central~\cite{REF:feng}.

More microscopically oriented approaches to the elastic
properties of vulcanized matter have also been made, in which
macromolecular degrees of freedom feature explicitly. Among these are
the ``phantom network''~\cite{REF:phantom} and ``affine
network''~\cite{REF:Florybook} approaches, as well as the comprehensive
discussion of rubber elasticity by Deam and Edwards~\cite{REF:DeamEd},
and others\cite{REF:Panyukov}.
These approaches focus on the well-cross-linked regime rather than the
lightly-cross-linked regime near the vulcanization
transition~\cite{REF:MarkErman}. 

Experimentally, the exponent $\shex$ has been addressed for several
systems (although mostly for gelation rather than
vulcanization): the results vary from $\shex\approx 2$~\cite{REF:exp2}
to $\shex \agt 3$~\cite{REF:exp3}. 
This wide discrepancy is not understood.

Stimulating though they certainly are, it must be recognized that
neither the classical~\cite{REF:Flory,REF:Stockmayer,REF:DobGor} nor
the percolation~\cite{REF:PGDGbook,REF:PGDGperc} approaches to the
physics of vulcanized matter explicitly include both crucial
ingredients: {\em thermal fluctuations\/} and {\em quenched
disorder\/}.  In recent years, an approach to the vulcanization
transition has been
developed~\cite{REF:prl_1987,REF:PMGandAZprl,REF:epl,REF:cross} that takes
into account both 
of these ingredients
in the context of a semi-microscopic model for flexible, randomly
cross-linked macromolecules.  This approach is very much inspired by
the work of Edwards and collaborators~\cite{REF:DeamEd,REF:Ball}, as
well as by concepts from the field of spin
glasses. Emerging from this more recent approach has been a detailed
picture of the {\em structure\/} of the amorphous solid state near to
the vulcanization transition, including, in particular, an explicit
form for the distribution of localization lengths.  What has not yet
been elucidated using this
approach is the second signature of the vulcanization transition,
namely the emergence of static {\em response\/} to shear deformations.
This issue is the focus of the present Letter.

\noindent
{\em Model\/}: At the heart of the theory of the {\em structure\/} of
the amorphous solid state\cite{REF:cross} is the analysis, employing
the techniques of replica statistical mechanics, of a semi-microscopic
model of $N$ macromolecules subject to random cross-linking
constraints.  This analysis leads to an order parameter
$\Omega_{\hat{k}}$ appropriate for diagnosing the amorphous solid
state, as well as a Landau free energy ${\cal
F}_{n}(\{\Omega_{\hat{k}}\})$ in terms of this order parameter.  A
detailed review of this theory has been given in
Ref.~\cite{REF:cross}.  The order parameter is defined via
\begin{equation}
\Omega_{\hat{k}}\equiv
\Big\langle
\frac{1}{N}\sum\nolimits_{i=1}^{N} \int_{0}^{1}\!\!\!\!ds
\exp\big(i{\hat{k}}\cdot{\hat c}_{i}(s)\big)
\Big\rangle_{n+1}^{\rm P}. 
\label{EQ:ReplicaOrder}
\end{equation}
Here, hatted vectors denote replicated collections of vectors, viz., 
${\hat{v}}\equiv\{{\bf v}^{0},{\bf v}^{1},\cdots,{\bf v}^{n}\}$,
their scalar product being 
${\hat{v}}\cdot{\hat{w}}\equiv
\sum_{\alpha=0}^{n}{\bf v}^{\alpha}\cdot{\bf w}^{\alpha}$, 
and the trajectories 
$\{\hat{c}(s)\}_{i=1}^{N}$ are the semi-microscopic configurations 
of the replicated macromolecules (where $0\le s\le 1$ is the arclength 
in units of the total arclength). 
$\langle\cdots\rangle_{n+1}^{\rm P}$ denotes an average for an
effective pure (i.e.\ disorder-free) system of $n+1$ coupled replicas of
the original system.  To model the disorder we make the
Deam-Edwards assumption~\cite{REF:DeamEd} that the statistics of the
cross-links is determined by the instantaneous correlations of the
uncrosslinked system.  This leads to the need to work with the $n\to 0$
limit of systems of $n+1$ (as opposed to $n$) replicas. The additional
replica, labeled by $\alpha=0$, represents the degrees of freedom of the
original system before cross-linking, or, equivalently,
describes the cross-link distribution. Consequently, any external strain
applied to the system {\em after} the permanent constraints have been created
will affect replicas $\alpha=1,\ldots,n$, but not replica
$\alpha=0$~\cite{REF:DeamEd}.  Thus, the order parameter
measures the correlations between the positions of individual particles
before and after the deformation is applied.

In the saddle-point approximation\cite{FNOTE:MF_DG}, the
disorder-aver\-aged free energy 
$f$ (per particle) in a $d$-dimensional system is obtained by
minimizing the replicated free-energy functional 
${\cal F}_{n}(\{\Omega_{\hat{k}}\})$~\cite{FNOTE:F_n_complete}:
\begin{equation}
f=d\lim_{n\rightarrow 0}\min_{\{\Omega_{\hat{k}}\}} 
{\cal F}_{n}\big(\{\Omega_{\hat{k}}\}\big).
\label{EQ:physical_f}
\end{equation}
As discussed in detail in Ref.~\cite{REF:cross}, the minimization in 
Eq.~(\ref{EQ:physical_f}) yields the liquid--amorphous-solid phase 
transition at a certain critical value of the cross-link density.  We
parametrize  
the excess cross-link density beyond this critical value by the control 
parameter $\epsilon$.  As the transition is continuous (i.e.\ near the
critical point the gel 
fraction is small and the typical localization length of localized 
particles is large), 
${\cal F}_{n}(\{\Omega_{\hat{k}}\})$ can be expanded in powers of 
the order parameter and gradients, with only low orders needing to
be retained~\cite{FNOTE:notmicro}:
\begin{eqnarray}
&&nd{\cal F}_{n}\big(\{\Omega_{\hat{k}}\}\big) = 
{\overline{\sum}}_{\hat{k}}
\big(-\epsilon+\frac{1}{2}|\hat{k}|^2\big)
\big\vert\Omega_{\hat{k}}\big\vert^{2}
\nonumber\\
&&\qquad\qquad
-\,{\overline{\sum}}_{{\hat{k}_1}{\hat{k}_2}{\hat{k}_3}}
\Omega_{\hat{k}_1}\,
\Omega_{\hat{k}_2}\,
\Omega_{\hat{k}_3}\,
\delta_{{\hat{k}_1}+{\hat{k}_2}+{\hat{k}_3}, {\hat{0}}}\,.
\label{EQ:LG_rescale}
\end{eqnarray}
The symbol ${\overline{\sum}}$ denotes a sum over replicated
wave-vectors that contain at least two nonzero component-vectors ${\bf
k}^{\alpha}$~\cite{FNOTE:omit}.  
The saddle-point equation for the free-energy functional near the
transition is exactly solved by the 
following hypothesis~\cite{REF:epl,REF:cross}: 
\begin{mathletters}
\begin{eqnarray}
\Omega_{\hat{k}} & = &
(1-q)\,\delta_{ {\hat{k}},\hat{0}} 
+q\,\delta_{\sum_{\alpha=0}^{n}{\bf k}^{\alpha},{\bf 0}}\,
\,\cpo^{\rm u}(\hat{k}),
\label{EQ:ord_par_hyp}
\\
\cpo^{\rm u}(\hat{k}) & \equiv & \int_{0}^{\infty}\!\!d\tau\,p(\tau)\,
{\rm e}^{-\hat{k}^{2}/2\tau}\,.
\label{EQ:cont_part_undef}
\end{eqnarray}
\end{mathletters}%
The physical motivation for this hypothesis comes from a picture in which 
a fraction $q$ of the monomers are localized around random mean positions 
${\bf b}_{i}(s)$ about which they execute harmonic thermal fluctuations 
over random  localization lengths $\xi_{i}(s)$.  Furthermore, the mean 
positions are assumed to he homogeneously distributed over the sample, 
and the localization lengths are characterized by the statistical 
distribution $2\xi^{-3} p(\xi^{-2})$.
Thus, delocalized and localized particles are, respectively, represented 
by the first and second terms on the RHS of Eq.~(\ref{EQ:ord_par_hyp}). The
$\delta$-factor in the second term 
comes from the homogeneity of the 
distribution of mean positions. The function $\cpo^{\rm u}(\hat{k})$,
which we refer to as the {\em continuous part} of the order parameter, encodes 
all the information about thermal fluctuations (the superscript u 
standing for ``unstrained'').  
The hypothesis~(\ref{EQ:ord_par_hyp}) and (\ref{EQ:cont_part_undef}) 
satisfies the saddle-point equations provided that~\cite{REF:epl,REF:cross} 
\begin{mathletters}
\begin{eqnarray}
0 & = & -2 q \epsilon + 3 q^{2},
\label{EQ:scqeq}
\\
\frac{\theta^{2}}{2} \frac{d\pi}{d\theta}
& = & (1-\theta)\,\pi(\theta)-
\int_{0}^{\theta} d\theta^{\prime}
\pi(\theta^{\prime})\pi(\theta-\theta^{\prime}), 
\label{EQ:scpieq}
\end{eqnarray}
\end{mathletters}%
where $\pi(\theta)$ is an $\epsilon$-independent scaling function 
such that $p(\xi^{-2})=(2/\epsilon)\,\pi(2/\epsilon\xi^2)$, and 
satisfies the boundary condition 
$\int_{0}^{\infty}\!\!\! d\theta \,\pi(\theta) =1$.  
Equation~(\ref{EQ:scqeq}) determines the localized fraction $q$: for
$\epsilon\le 0$ we obtain $q=0$ (i.e.\ the liquid phase, which has a
vanishing static shear modulus); for $\epsilon>0$ we obtain 
$q=2\epsilon/3$, corresponding to the amorphous solid state,
which is the state on which we shall focus from now on. 

\noindent
{\em Response to shear strain\/}: We now set about determining the 
free-energy cost associated with making static shear deformations of 
the system.  To do this, we consider the effect of changing the shape 
of the container (on which we have imposed periodic boundary 
conditions).  We characterize the deformation by the ($d\times d$) 
matrix ${\fft S}$, which describes the change in position of any point
${\bf b}$ at the boundary of the system as follows:
${\bf b}\rightarrow{\fft S}\cdot{\bf b}$. 
For example, for $d=3$ and for a 
deformation in which the $x$, $y$ and $z$ cartesian components of the
position vector are, respectively, elongated by the factors
$\lambda_{x}$, $\lambda_{y}$  
and $\lambda_{z}$, the matrix ${\fft S}$ has the form 
${\rm diag}(\lambda_{x},\lambda_{y},\lambda_{z})$. 
As we are concerned with the free-energy cost of pure shear strains, 
we shall assume that the deformation leaves the volume $V$ of the system 
unchanged, i.e., ${\rm Det}\,{\fft S}=1$. 
For considering infinitesimal strains, it is convenient to define the 
(symmetric) strain tensor 
${\fft J}\equiv\frac{1}{2}({\fft S}+{\fft S}^{\rm T})-{\fft I}$. 
Here ${\fft S}^{\rm T}$ is the transpose of ${\fft S}$, and ${\fft I}$ 
is the identity matrix. As ${\rm Det}\,{{\fft S}}=1$ we have 
${\rm Tr}\,{\fft J}=0$, to first order in the deformation. 

Before taking the thermodynamic limit, the system is finite in extent,
so that the 
Fourier representation of any function of position consists of a
superposition of plane waves with discrete wave-vectors.  In particular,
the order parameter (which is a function on replicated Fourier space) is
only defined at a discrete set of points.  Now, under strain the
boundaries in position space are displaced and, as a consequence, the
discretization in replicated Fourier space changes.  As mentioned above,
any external strain applied to the system after the permanent
constraints have been created will affect replicas $\alpha=1,\ldots,n$,
but not replica $\alpha=0$~\cite{REF:DeamEd}.  Therefore, the change in
the discretization of the wave vectors occurs only for 
$\alpha=1,\ldots,n$, but not  $\alpha=0$.  For convenience, 
we shall use the symbols $R^{\rm u}$ and $R^{\rm s}$ to denote the sets 
of allowed replicated wave-vectors corresponding, respectively, to the 
unstrained and strained systems.

Conceptually, there are two sources for the change in free
energy [Eq.~(\ref{EQ:physical_f})] under deformation: the change in
the expression for the free energy functional itself, and the consequent
change in the value of the order parameter that solves the
saddle-point equation.  The free-energy functional for the strained
system ${\cal F}_{n}^{\rm s}(\{\Omega_{\hat{k}}\})$ is obtained by repeating, 
step-by-step, the construction of the free-energy functional for the 
unstrained system ${\cal F}_{n}(\{\Omega_{\hat{k}}\})$.  The 
result~\cite{FNOTE:construction} is that the coefficient in front
of each term is unaltered, the only change being the replacement of
each sum $\sum_{\hat{k}}$ over the old set of discrete replicated 
wave-vectors (i.e.\ $\hat{k}\in R^{\rm u}$) by a sum over the 
new set of discrete replicated wavevectors $\sum_{\hat{k}\in R^{\rm s}}$.
As a result, the saddle-point equation for the strained system becomes
\begin{equation}
 0 = 
 2\big(-\epsilon+\frac{1}{2}|{\hat{k}}|^2\big)
 \Omega_{\hat{k}} 
  -3{{\displaystyle{\overline\sum}}
	\atop{\scriptstyle\hat{k}_1\hat{k}_2\in R^{\rm s} }}
 \Omega_{\hat{k}_1}\,
 \Omega_{\hat{k}_2}\,
 \delta_{{\hat{k}_1}+{\hat{k}_2},{\hat{k}}}\,.
 \label{EQ:LG_saddle}
\end{equation}

We now obtain the order parameter for the strained system by making a
physically motivated hypothesis similar to the one made for the 
unstrained system.  First, for each localized monomer in the unstrained 
system we envision that its old mean position ${\bf b}_{i}(s)$ is 
displaced to a new mean position 
${\fft S}\cdot{\bf b}_{i}(s) + {\bf r}_{i}(s)$, where 
${\fft S}\cdot{\bf b}_{i}(s)$ is the affine displacement of the old 
position~\cite{REF:Florybook} and ${\bf r}_{i}(s)$ is a random 
additional displacement, which we take to be uncorrelated with 
${\bf b}_{i}(s)$.  With the assumption that (as in the unstrained system) 
there is no correlation between the extent (including shape) of the thermal
fluctuations of a monomer about its mean position and the mean position
itself, we arrive at the hypothesis
\begin{equation}
\Omega_{\hat{k}}= (1-q)\,\delta_{ {\hat{k}},\hat{0}}
+q\,\delta_{{\bf k}^{0}
	+{\fft S}^{\rm T}\cdot\sum_{\alpha=1}^{n}{\bf k}^{\alpha},
	{\bf 0}}\,
\, \cpo^{\rm s}(\hat{k}), 
\label{EQ:ord_par_hyp_s} 
\end{equation}
where $\cpo^{\rm s}(\hat{k})$ is the continuous part of the order 
parameter in the strained system.  Now, to construct a form for 
$\cpo^{\rm s}(\hat{k})$ we consider a conjecture for the form of 
$\langle e^{i{\bf k}\cdot{\bf c}_{i}(s)}\rangle^{\rm s}_{\chi}$ (i.e.\
the thermal expectation values of the Fourier-transformed individual 
particle densities in the 
strained (s) 
system for a specific disorder realization $\chi$):  
\begin{eqnarray}
&& \exp\big( i{\bf k}\!\cdot\! \left\{ {\fft S}\!\cdot\!{\bf b}_{i}(s)
+ {\bf r}_{i}(s) \right\} \big)
\nonumber \\
&& \qquad\qquad 
 \times \exp\big(-\xi^{2}_{i}(s)\,{{\bf k}\!\cdot\! \left\{ {\fft I}
		+\eta_{i}(s) \, {\fft J} \right\} \!\cdot\!{\bf k}}/2\big).
\end{eqnarray}
We expect the gaussian 
probability cloud to be isotropic, except for a correction 
due to the distortion.  For infinitesimal distortions, 
this correction should be proportional to ${\fft J}$ and have a random 
magnitude $\eta_{i}(s)$.  For example, if $\eta_{i}(s)=2$ then the 
probability cloud would have been affinely distorted.  On the other hand, 
if $\eta_{i}(s)=0$ then the probability cloud would remain spherical. 
Assuming, further, that the random displacement ${\bf r}_{i}(s)$ also has a
probability distribution shaped by a combination of ${\fft I}$ and
${\fft J}$, and expanding to lowest nontrivial order in the
deformation~\cite{FNOTE:order}, we obtain the hypothesis:
\begin{equation}
\cpo^{\rm s}(\hat{k})\!=\!q\!\int_{0}^{\infty}\!\!\!d\theta\,
e^{-\hat{k}^2/{\epsilon \theta}}
\Big(\pi(\theta) -
{\wimpf(\theta)\over{\epsilon}}
\sum_{\alpha=1}^{n}
{\bf k}^{\alpha}\!\cdot\!{\fft J}\!\cdot\!{\bf k}^{\alpha}\Big).
\label{EQ:def_hyp}
\end{equation}
Here, $\wimpf(\theta)$ is a second scaling function, which describes the
change in the continuous part of the order parameter due to the
deformation.

Alternative motivation for the form of $\cpo^{\rm s}(\hat{k})$ 
runs as follows. Let us assume that for small strains 
$\cpo^{\rm s}(\hat{k})$ is unchanged
under simultaneous rotations of ${\fft S}$ and $\hat{k}$. 
As it is only a function of $\hat{k}^{2}$ this property 
certainly holds for $\cpo^{\rm u}(\hat{k})$, and it therefore 
also holds for the difference between $\cpo^{\rm s}(\hat{k})$ and 
$\cpo^{\rm u}(\hat{k})$. 
To first order in ${\fft J}$ this difference can only contain the 
following terms: 
(i)~a linear function of 
$\sum_{\alpha=1}^{n}{\bf k}^{\alpha}
	\cdot{\fft J}\cdot{\bf k}^{\alpha}$
and (ii)~a product of an invariant linear function of ${\fft J}$ 
with an invariant function of $\hat{k}$.  
The only quantity linear in ${\fft J}$ and invariant under 
rotations is ${\rm Tr}\,{\fft J}$,
which is zero for infinitesimal shear strains, as mentioned above. 
Thus we recover Eq.~(\ref{EQ:def_hyp}).

By inserting the hypothesis given by Eqs.~(\ref{EQ:ord_par_hyp_s}) and
(\ref{EQ:def_hyp}) into the saddle-point condition~(\ref{EQ:LG_saddle}), 
we recover Eqs.~(\ref{EQ:scqeq}) and (\ref{EQ:scpieq}) for $q$ and 
$\pi(\theta)$, together with the 
condition: 
\begin{equation}
\frac{\theta^{2}}{2}\frac{d\wimpf}{d\theta}
\!=\!
(1-\theta)\,\wimpf(\theta)
\!-\!
\frac{2}{\theta^{2}}\int_{0}^{\theta}\!\!\!d\theta^{\prime}\,
{\theta'}^{2}\,\wimpf(\theta^{\prime})\,\pi(\theta-\theta^{\prime}).
\label{EQ:scphieq}
\end{equation} 
The boundary condition $\lim_{\theta \to \infty} \theta^{2}
\wimpf(\theta) = 0$ 
stems from the fact that, by Eq.~(\ref{EQ:ReplicaOrder}),
$\lim_{|{\hat{k}}| \to \infty} {\Omega_{\hat{k}}} = 0$. 
The only solution of
Eq.~(\ref{EQ:scphieq}) that satisfies the boundary condition is the null 
function: $\wimpf(\theta)\equiv 0$~\cite{FNOTE:onlysol}.
This result implies the first (and, {\it a priori\/}, the most 
surprising) result of this Letter: the continuous part of the order 
parameter {\em does not change\/} to first order in the strain, 
i.e., $\cpo^{\rm s}(\hat{k})=\cpo^{\rm u}(\hat{k})$. 
This conclusion is consistent with the phantom network 
picture~\cite{REF:phantom,REF:MarkErman}. 
It also suggests that  
$\cpo^{\rm s}(\hat{k})=\cpo^{\rm u}(\hat{k})$ for finite 
(and not merely infinitesimal) deformations, and indeed the resulting
order-parameter hypothesis turns out to satisfy the saddle-point
equation for arbitrarily strained systems.

We now have all the ingredients necessary to compute the 
change in the free energy, to leading order in $\epsilon$,
due to the deformation of the system:
\[
\Delta f\!=\!
d\lim_{n\rightarrow 0} 
\left[
{\cal F}^{\rm s}_{n}\big(\{\Omega^{\rm s}_{\hat{k}}\}\big)
\!-\!
{\cal F}_{n} \big(\{\Omega^{\rm u}_{\hat{k}}\}\big)\right]
\!=\!
\frac{2\epsilon^{3}}{27}{\rm Tr}\,
({\fft S}\cdot{\fft S}^{\rm T} - {\fft I}),
\]
where $\Omega^{\rm s}_{\hat{k}}$ and $\Omega^{\rm u}_{\hat{k}}$ are,
respectively, the saddle-point values of the order parameter for the
strained and unstrained systems. Thus we can extract the value of the
static shear modulus for the amorphous solid state near the
solidification transition (with physical units restored):
$\shmd=k_{\rm B}TNC\epsilon^{3}$, where $k_{\rm B}$ is Boltzmann's
constant, $T$ is the temperature, and $C$ is a model-dependent
positive constant.  Hence we see that the static shear modulus near
the vulcanization transition is characterized by the exponent
$\shex=3$, in agreement with the classical
result~\cite{REF:DobGor,REF:PGDGbook}.  A simple scaling argument,
viz., that the modulus should scale as two powers of the order
parameter ($q^2$) and two powers of the gradient ($\xi^{-2}_{\rm
typ}$), leads to the same value for $\shex$.

\noindent 
{\em Concluding remarks\/}: We have presented a microscopic
derivation of the static elastic response of a system of randomly
cross-linked macromolecules near the amorphous solidification
transition.  In the picture that emerges, it is seen: (i)~that the
amorphous solid state, which was previously shown to be characterized
structurally 
by the localization of a nonzero fraction of particles, is also
characterized by having a nonzero static shear modulus; (ii)~that the
static shear modulus scales as the third power of the excess
cross-link density (beyond its value at the transition)
\cite{FNOTE:previous}; and (iii)~that the form of localization
exhibited by the particles is left unchanged by the strain.  
It is, however, not implausible that strain-induced changes would
emerge from a more detailed analysis of the effects of the
excluded-volume interaction, at least at higher crosslink densities.
Being dependent only on the form of the free-energy
functional~\cite{REF:landau,FNOTE:notmicro}, and not any specific
semi-microscopic model, the approach to elasticity described here
should be generally applicable not only to systems of randomly
cross-linked flexible macromolecules, but also to other equilibrium
amorphous solid forming systems.

\noindent 
{\em Acknowledgments\/}: We thank 
S.~Barsky, B.~Jo{\'o}s, M.~Plischke and A.~Zippelius
for stimulating discussions. We gratefully acknowledge support
from National Science Foundation grant DMR94-24511.
%----------

\end{multicols}

\begin{references}
\bibitem{FNOTE:MF_DG}
In the vulcanization of polymer melts, the width of the critical
region, in which mean-field theory fails,
vanishes in the limit of long macromolecules for $d \ge 3$\cite{REF:PGDGbook}.
\bibitem{FNOTE:F_n_complete} 
The complete expression for the free energy functional for any
cross-link density is given in Eq.~(5.12b) of Ref.~\cite{REF:cross}.
\bibitem{FNOTE:notmicro}
This form for ${\cal F}_{n}(\{\Omega_{\hat{k}}\})$ can be
obtained either from a semi-microscopic model, as envisioned here, or
via an argument involving symmetries and the continuity of the
transition in the context of a Landau theory~\cite{REF:landau}.
\bibitem{FNOTE:order}
The shear modulus is
determined by an expansion of the free energy to quadratic order in
the deformation.
Thus, only terms to linear order are
needed in $\cpo^{\rm s}$.
\end{references}
\end{document}